\documentclass[a4page, 10pt]{article}
\usepackage{amssymb,amsmath}
\usepackage{graphicx}
\numberwithin{equation}{section}

\begin{document}

\title{Quantum Gravity in Flat Spacetime}

\author{Jarmo M\"akel\"a\footnote{Vaasa University of Applied Sciences,
Wolffintie 30, 65200 Vaasa, Finland, email: jarmo.makela@vamk.fi}}

\maketitle

\begin{abstract}

Inspired by Einstein's Strong Principle of Equivalence we consider the effects of quantum mechanics to the gravity-like phenomena experienced by an observer in a uniformly accelerating motion in flat spacetime. Among other things, our model of quantum gravity, derived from the first principles, predicts the Unruh effect, and a discrete area spectrum for spacelike two-surfaces.
 
\end{abstract}

\maketitle

\section{Introduction}

When Albert Einstein began to formulate his general theory of relativity, he used as a starting point his Strong Principle of Equivalence. In broad terms, this principle states that it is impossible to decide, by means of local measurements, whether one is in a gravitational field or in an accelerating frame of reference. In this sense the effects of gravity are equivalent with those of acceleration. \cite{yy, kaa} For instance, if an observer finds all bodies in his surroundings to fall with the same acceleration, the observer has no way to tell, whether he is in a uniform gravitational field, caused by the curvature of spacetime, or in a uniformly accelerating motion in flat spacetime. 

 One of the greatest challenges of modern physics is to create the quantum theory of gravity, which is supposed to bring together general relativity and quantum mechanics. It appears to the author that in the attempts to quantize gravity developed so far the Equivalence Principle has received insufficient attention. It is possible that we should take the Equivalence Principle as the starting point of quantum gravity in the same way as Einstein took it as the starting point of his general relativity. 

    Inspired by these thoughts we shall, in this paper, quantize the effects of gravity in {\it flat spacetime} from the point of view of an observer in a uniformly accelerating motion. According to the Equivalence Principle quantization of gravity in curved spacetime should, at least locally, be equivalent with the quantization of gravity in flat spacetime from the point of view of an acccelerating observer. We shall see that the resulting "quantum gravity in flat spacetime" may indeed be developed systematically, beginning from the first principles. 

In our approach we consider the gravity-like effects on a spacelike two-plane at rest with respect to our accelerating observer, and it plays a role somewhat similar to that of the event horizon of a black hole.  The standard rules of quantum mechanics applied to our model imply that the accelerating plane has a discrete structure in the sense that it consists of a finite number of separate constituents, all of them having an equally spaced area spectrum. It also predicts the Unruh effect, according to which an observer in an accelerating motion detects thermal radiation with a characteristic temperature, which is proportional to the proper acceleration of the observer. 

Unless otherwise stated, we shall always use the natural units, where $\hbar = G = c = k_B=1$.

\section{Action}

Whenever we attempt to quantize gravity, we may use the classical Einstein-Hilbert action
\begin{equation}
S_{EH} := \frac{1}{16\pi}\int_M R\sqrt{-g}\,d^4x
\end{equation} 
as the starting point. In this equation $R$ is the Riemann curvature scalar, and $g$ is the determinant of the metric tensor of spacetime. We have integrated over the whole spacetime $M$. However, if the spacetime region under consideration possesses a {\it boundary}, we must supplement the Einstein-Hilbert action with the {\it Gibbons-Hawking boundary term} \cite{koo}
\begin{equation}
S_{GH} := \frac{1}{8\pi}\int_{\partial M} K\,dV,
\end{equation}
where $K$ is the trace of the exterior curvature tensor induced on the boundary $\partial M$ of spacetime. $dV$ is the volume element on the boundary, and we have integrated over the whole boundary. So the whole action takes the form:
\begin{equation}
S = S_{EH} + S_{GH} = \frac{1}{16\pi}\int_M R\sqrt{-g}\,d^4x + \frac{1}{8\pi}\int_{\partial M} K\,dV.
\end{equation}

   In flat spacetime the Riemann curvature scalar $R\equiv 0$, and we are left with the Gibbons-Hawking boundary term only. Hence the action becomes to:
\begin{equation}
S = \frac{1}{8\pi}\int_{\partial M} K\,dV.
\end{equation}
Even if spacetime were flat, the action may still be non-zero, if the boundary of spacetime has been chosen appropriately. Because of the presence of the potentially non-zero boundary term, the quantum-mechanical properties of the gravity-like effects caused by a choice of a non-inertial frame of reference are far from trivial. 
           
    In this paper we first pick up from the four-dimensional flat spacetime an inertial frame of reference, where each point of spacetime is identified by means of four coordinates $(T, X, Y, Z)$. The coordinate $T$ is timelike, and the coordinates $X$, $Y$ and $Z$ spacelike. When written in terms of these coordinates the line element of spacetime takes the flat Minkowski form:
\begin{equation}
ds^2 = -dT^2 + dX^2 + dY^2 + dZ^2.
\end{equation}
We choose the boundary of spacetime to consist of five timelike hyperplanes, where $X=X_1$, $Y=Y_1$, $Y=Y_2$, $Z=Z_1$ and $Z=Z_2$ such that $Y_2>Y_1$ and $Z_2>Z_1$. In other words, the boundary of spacetime is a three-dimensional, rectangular {\it box}, which proceeds in time, and where the faces stay still in our frame of reference. 

      One of the faces of the box, however, is not assumed to stay still, but it is assumed to {\it accelerate} with constant proper acceleration $a$ to the direction of the positive $X$-axis. As a consequence, the $X$-coordinates of the ponts of the face satisfy an equation:
\begin{equation}
X^2 - T^2 = \frac{1}{a^2},
\end{equation}
which implies:
\begin{equation}
X = \sqrt{T^2 + \frac{1}{a^2}}.
\end{equation}
So the boundary of spacetime consists, as a whole, of five timelike hyperplanes, and of a one timelike hypersurface, which is created, when a spacelike two-plane accelerates along the $X$-axis.

        The exterior curvature of all of the timelike hyperplanes is zero, and so they do not contribute to the action. The exterior curvature of the hypersurface created by the accelerating plane, however, is non-zero. To calculate the trace of the exterior curvature tensor on that hypersurface we introduce the {\it Rindler coordinates} to our spacetime. As it is well known, between the Rindler coordinates $(t,x)$ and the coordinates $(T,X)$ there is the relationship: \cite{nee}
\begin{subequations}
\begin{eqnarray}
t &=& \tanh^{-1}\left(\frac{T}{X}\right),\\
x &=& \sqrt{X^2 - T^2}.
\end{eqnarray}
\end{subequations}
When written in terms of the Rindler coordinates, together with the coordinates $Y$  and $Z$, the line element of spacetime takes the form:
\begin{equation}
ds^2 = -x^2\,dt^2 + dx^2 + dY^2 + dZ^2.
\end{equation}
An observer with constant Rindler coordinate $x$, together with constant coordinates $Y$ and $Z$, has constant proper acceleration $a$, which is related to $x$ such that
\begin{equation}
x = \frac{1}{a},
\end{equation}
which follows from Eqs. (2.6) and (2.8b). Eq. (2.8 a), in turn, implies that the Rindler time coordinate $t$ gives the {\it boost angle} $\phi$ of an observer with constant proper acceleration. 

     On the timelike hypersurface created, when a plane is in an accelerating motion with constant proper acceleration $a$ the Rindler coordinate $x$ is a constant. The only non-zero component of the exterior curvature tensor on that hypersurface is:
\begin{equation}
K_{tt} = \Gamma_{tt}^x = x,
\end{equation}
and therefore its trace is:
\begin{equation}
K = g^{tt}K_{tt} = -\frac{1}{x}.
\end{equation}
The volume element on our hypersurface is:
\begin{equation}
dV = x\,dY\,dZ\,dt,
\end{equation}
and so we find that the action in Eq. (2.7) takes, if the Rindler time coordinate $t$ lies within an interval $[t_i,t_f]$, the form:
\begin{equation}
S = -\frac{1}{8\pi}A(t_f - t_i),
\end{equation}
where $A$ is the area of the accelerating plane. \cite{vii}

\section{Hamiltonian}

The bridge from the action to the Hamiltonian $H(q_j,p_j;t)$ of any system is formed by the Hamilton-Jacobi equation:
\begin{equation}
H(q_j,\frac{\partial S}{\partial q_j};t) + \frac{\partial S}{\partial t} = 0,
\end{equation}
where $S$ is the principal function of the system. The quantities $q_j$ are the coordinates of the configuration space, and the quantities $p_j = \frac{\partial S}{\partial q_j}$ the corresponding coordinates of the momentum space. The principal function of the system has been denoted by the same symbol as its action for a very good reason: As it is well known, the principal function agrees with the action integral calculated along the classical path. Hence it follows that if we know the action $S$ calculated along the classical path, the Hamiltonian of the system will be: \cite{kuu}
\begin{equation}
H = -\frac{\partial S}{\partial t}.
\end{equation}

    Our derivation of the classical Hamiltonian of the gravitational field is based on Eq. (3.2). Before proceeding to the use of Eq. (3.2), however, we must decide, which time coordinate to use. The very idea of this paper is to consider the effects of gravity from the point of view of a uniformly accelerating observer. It is therefore sensible to take the {\it proper time} $\tau$ of the observer as the time coordinate. Eq. (2.9) implies that between the proper time $\tau$ and the Rindler time $t$ there is the relationship:
\begin{equation}
d\tau = x\,dt.
\end{equation}
Writing
\begin{equation}
H = -\frac{\partial S}{\partial\tau},
\end{equation}
and using Eqs. (2.10) and (2.14) we therefore find:
\begin{equation}
H = \frac{a}{8\pi}A.
\end{equation}

    Eq. (3.5) gives the classical Hamiltonian of the gravitational field from the point of view of an observer moving along a plane with area $A$ and proper acceleration $a$. Remarkably, the Hamiltonian in Eq. (3.5) is identical to the one obtained in several papers in spacetimes with {\it horizon}, from the point of view of an observer with constant proper acceleration $a$, just outside of a horizon with area $A$. [7-14] When the proper acceleration $a$ of the observer tends to infinity, Eq. (2.10) implies that the Rindler coordinate $x$ tends to zero, which means that in this limit our plane tends to a Rindler horizon of flat spacetime. 

    We may always define a diffeoforphism $f$ from the unit square $I^2:= [0,1]\times [0,1]$ to the accelerating plane, which we shall henceforth denote by $B$. In general, this map $f:I^2\longrightarrow B$ is of the form:
\begin{equation}
f(\chi^1,\chi^2) = (Y(\chi^1,\chi^2), Z(\chi^1,\chi^2))
\end{equation}
for all $\chi^1, \chi^2\in [0,1]$, where the coordinates $Y(\chi^1,\chi^2)$ and $Z(\chi^1,\chi^2)$ of the point $(Y,Z)$ on the plane $B$ are functions of the coordinates $\chi^1$ and $\chi^2$ of the point $\chi = (\chi^1, \chi^2)$ in $I^2$. Defining the function
\begin{equation}
p(\chi) := p(\chi^1,\chi^2) := \frac{1}{8\pi}\left\vert\begin{matrix}
                                                                              \frac{\partial Y(\chi^1,\chi^2)}{\partial \chi^1}&\frac{\partial Y(\chi^1,\chi^2)}{\partial\chi^2}\\
\frac{\partial Z(\chi^1,\chi^2)}{\partial\chi^1}&\frac{\partial Z(\chi^1,\chi^2)}{\partial\chi^2}\end{matrix}\right\vert
\end{equation}
we may write the area of the plane $B$ as:
\begin{equation}
A = 8\pi\int_{I^2}p(\chi)\,d^2\chi,
\end{equation}
where we have integrated over the unit square. As a consequence, the Hamiltonian of the gravitational field takes, from the point of view of our accelerating observer, the form:
\begin{equation}
H = a\int_{I^2}p(\chi)\,d^2\chi.
\end{equation}
It must be emphasized that the proper acceleration $a$ is {\it not} a dynamical variable of the system, but just a {\it parameter}, which determines the observer. 

   We shall now take the quantities $p(\chi)$ associated with the points $\chi$ as the coordinates of the momentum space. The corresponding coordinates of the configuration space we shall denote by $q(\chi)$. They obey the Hamiltonian equations of motion:
\begin{equation}
\dot{q}(\chi) = \frac{\delta H}{\delta p(\chi)} = a,
\end{equation}
where the dot means the derivative with respect to the proper time $\tau$ of the accelerating observer. The general solution to Eq. (3.10) is:
\begin{equation}
q(\chi) = a\tau + C(\chi),
\end{equation}
where $C(\chi)$ is an arbitrary constant function of the proper time $\tau$, and depends on the coordinates $(\chi^1,\chi^2)$ of the point $\chi$ only. Eqs. (2.10) and (3.3) imply that the quantity $a\tau$ may be identified as the {\it boost angle} $\phi$ of the observer. Hence we may write Eq. (3.11) as:
\begin{equation}
q(\chi) = \phi + C(\chi).
\end{equation}
Since our Hamiltonian in Eq. (3.10) does not involve $q(\chi)$, the Hamiltonian equations of motion written for the variables $p(\chi)$ read as:
\begin{equation}
\dot{p}(\chi) = -\frac{\delta H}{\delta q(\chi)} = 0,
\end{equation}
which means that the quantities $p(\chi)$ are constants of motion of the system.

\section{Quantization}

When we go over from the classical to the quantum-mechanical consideration of the  gravitational field in flat spacetime we must replace the classical quantities $q(\chi)$ and $p(\chi)$ by the corresponding operators $\hat{q}(\chi)$ and $\hat{p}(\chi)$ obeying the canonical commutation relations:
\begin{subequations}
\begin{eqnarray}
\lbrack\hat{q}(\chi),\hat{q}(\chi')\rbrack = \lbrack\hat{p}(\chi),\hat{p}(\chi')\rbrack &=& 0,\\
\lbrack\hat{q}(\chi),\hat{p}(\chi')\rbrack &=& i\delta^2(\chi,\chi')
\end{eqnarray}
\end{subequations}
at all points $\chi, \chi'\in I^2$. In Eq. (4.1b) $\delta^2(\chi,\chi')$ is the two-dimensional delta function in $I^2$. When written in terms of these operators the classical Hamltonian $H$ in Eq. (3.9) becomes replaced by the corresponding Hamiltonian operator
\begin{equation}
\hat{H} = a\int_{I^2}\hat{p}(\chi)\,d^2\chi,
\end{equation}
and therefore the time-independent Schr\"odinger equation
\begin{equation}
\hat{H}\vert\psi\rangle = E\vert\psi\rangle
\end{equation}
written for the energy eigenstates $\vert\psi\rangle$ of the gravitational field takes the form:
\begin{equation}
a\int_{I^2}\hat{p}(\chi)\vert\psi\rangle = E\vert\psi\rangle.
\end{equation}
Identifying the energy $E$ of the gravitational field with the classical Hamiltonian $H$ in Eq. (3.9) we observe that this equation reduces to:
\begin{equation}
\int_{I^2}\hat{p}(\chi)\vert\psi\rangle = \int_{I^2}p(\chi)\vert\psi\rangle.
\end{equation}

   Eq. (4.5) has an obvious solution, where $\vert\psi\rangle$ is the common eigenstate of all of the operators $\hat{p}(\chi)$, which means that:
\begin{equation}
\hat{p}(\chi)\vert\psi\rangle = p(\chi)\vert\psi\rangle
\end{equation}
for all $\chi\in I^2$. If the eigenstate $\vert\psi\rangle$ is normed to unity, we must have:
\begin{equation}
\langle\psi\vert\psi\rangle = 1.
\end{equation}
We shall take Eqs. (4.6) and (4.7) as the {\it postulates} of our model, and see, where this will take us.

   In the {\it position representation} the energy eigenstates $\vert\psi\rangle$ are expressed as functionals $\psi[q(\chi)]$ of the functions $q(\chi)$, and they may be regarded as the wave functions of the gravitational field. The inner product between the wave functions $\psi_1[q(\chi)]$ and $\psi_2[q(\chi)]$ is defined as:
\begin{equation}
\langle\psi_1\vert\psi_2\rangle := \int\psi_1^*[q(\chi)]\psi_2[q(\chi)]\,\mathcal{D}[q(\chi)],
\end{equation}
where $\mathcal{D}[q(\chi)]$ is an appropriate integration measure in the space of the functions $q(\chi)$. The momentum operators $\hat{p}(\chi)$ are expressed as functional differential operators:
\begin{equation}
\hat{p}(\chi) := -i\frac{\delta}{\delta q(\chi)},
\end{equation}
and so Eq. (4.6) takes the form:
\begin{equation}
-i\frac{\delta\psi[q(\chi)]}{\delta q(\chi)} = p(\chi)\psi[q(\chi)].
\end{equation}
Fortunately, this functional deifferential equation may be solved explicitly. Its general solution is:
\begin{equation}
\psi[q(\chi)] = \mathcal{N}\exp\left[i\int_{I^2}p(\chi)q(\chi)\,d^2\chi\right],
\end{equation}
where $\mathcal{N}$ is an arbitrary complex number.

   As one may observe, the functional $\psi[q(\chi)]$ in Eq. (4.11) is uniquely determined, up to the constant coefficient $\mathcal{N}$, by the function $p(\chi)$, which gives, at each point $\chi$, the eigenvalue of the operator $\hat{p}(\chi)$. If we denote by $p'(\chi)$ another eigenvalue of $\hat{p}(\chi)$, the corresponding eigenfunction is:
\begin{equation}
\psi'[q(\chi)] = \mathcal{N}'\exp\left[i\int_{I^2}p'(\chi)q(\chi)\,d^2\chi\right],
\end{equation}
and since eigenstates associated with different eigenvalues must be orthogonal, we must have:
\begin{equation}
\int\psi^*[q(\chi)]\psi'[q(\chi)]\,\mathcal{D}[q(\chi)] = 1,
\end{equation}
whenever
\begin{equation}
p'(\chi)\equiv p(\chi),
\end{equation}
and
\begin{equation}
\int\psi^*[q(\chi)]\psi'[q(\chi)]\,\mathcal{D}[q(\chi)] = 0,
\end{equation}
if $p'(\chi)$ differs from $p(\chi)$ at any point $\chi$. In Eq. (4.12) $\mathcal{N}'$ is an arbitrary complex number such that Eq. (4.13) is satisfied.

   We now divide the unit square $I^2$ in $N$ non-intersecting subsets $D_j$ $(j=1,2,\dots,N)$ such that $D_j\cap D_k=\emptyset$ for all $j \ne k$, where $j, k \in\lbrace 1,2,\dots,N\rbrace$, and
\begin{equation}
I^2 = D_1\cup D_2\cup\cdots\cup D_N.
\end{equation}
We shall assume that each subset $D_j$ has the same area
\begin{equation}
\Delta\chi := \frac{1}{N}.
\end{equation}
We shall also assume that in the limit, where $N$ tends to infinity, the supremum of the diameters of the subsets $D_j$ tends to zero. Picking up from each of the subsets $D_j$ a point $\chi_j$ we may write the integral in the exponential in Eq. (4.11) as a Riemann integral:
\begin{equation}
\int_{I^2}p(\chi)q(\chi)\,d^2\chi = \lim_{N\rightarrow\infty}\left(\sum_{j=1}^Np_jq_j\Delta\chi\right),
\end{equation}
where we have denoted:
\begin{subequations}
\begin{eqnarray}
p_j &:=& p(\chi_j),\\
q_j &:=& q(\chi_j)
\end{eqnarray}
\end{subequations}
for all $j=1,2,\dots,N$. Hence we have:
\begin{equation}
\psi[q(\chi)] = \mathcal{N}\lim_{N\rightarrow\infty}\left\lbrack\prod_{j=1}^N\exp(ip_jq_j\Delta\chi)\right\rbrack.
\end{equation}

   Eq. (4.20) enables us to define the integral on the left hand side of Eq. (4.13):
\begin{equation}
\int\psi^*[q(\chi)]\psi'[q(\chi)]\,\mathcal{D}[q(\chi)] := \mathcal{N}^*\mathcal{N}'\lim_{N\rightarrow\infty}\left\lbrace\prod_{j=1}^N\int_{-L/2}^{L/2}\exp[i(p'_j - p_j)q_j\,\Delta\chi]\,dq_j\right\rbrace,
\end{equation}
where we have denoted ${p'}_j:=p'(\chi_j)$. As one may see, we have introduced a new, positive number $L$. We might, of course, have taken $L$ to infinity, and thereby integrate $q_j$ from the negative to the positive infinity. Such a choice, however, would make the integral in Eq. (4.21) infinite. Because of that we shall henceforth take $L$ to be a {\it fixed number}. If we want Eqs. (4.13) and (4.15) to be satisfied, we must have, for every $j=1,2,\dots,N$:
\begin{equation}
p_j\Delta\chi = n_j\frac{2\pi}{L},
\end{equation}
where $n_j = 0,\pm 1,\pm 2\,\dots$. 

   The diffeomorphism $f:I^2\longrightarrow B$ defined in Eq. (3.6) maps every subset $D_j$ of $I^2$ to a certain piece, or constituent 
\begin{equation}
U_j = f(D_j)
\end{equation}
of the accelerating plane $B$. These constituents have the properties:
\begin{equation}
U_j\cap U_k = \emptyset
\end{equation}
for all $j \ne k$, and
\begin{equation}
B = U_1\cup U_2\cup\cdots\cup U_N.
\end{equation}
In the limit, where the number $N$ of the constituents becomes very large, we may view the quantity
\begin{equation}
A_j := 8\pi p_j\Delta\chi
\end{equation}
as the area of the constituent $j$. According to Eq. (4.22) the possible area eigenvalues of the constituent are:
\begin{equation}
A_j = n_j\frac{16\pi^2}{L}.
\end{equation}
Obviously, negative area eigenvalues do not make sense, and therefore we shall ignore the negative values of the quantum numbers $n_j$, and keep only the non-negative ones. In other words, $n_j = 0, 1, 2,\dots$ in Eq. (4.27). 

   As one may observe from Eq. (4.27), the areas of the constituents of the accelerating plane $B$ have a {\it discrete spectrum} with an equal spacing. Unless the plane has an infinite number of constituents with zero area, we are therefore forced to conclude that the number of the constituents of the plane is {\it finite}. This number we shall denote by $N$. The total area of the plane is the sum of the areas of its constituents:
\begin{equation}
A = A_1 + A_2 + \cdots + A_N.
\end{equation}
So we see that the postulates (4.6) and (4.7) of our model imply that our plane necessarily has a {\it discrete structure} in the sense that it consists of a finite number of separate constituents. The possible area eigenvalues of the plane are of the form:
\begin{equation}
A = \frac{16\pi^2}{L}(n_1 + n_2 +\cdots + n_N).
\end{equation}
The eigenfunctions associated with these area eigenvalues are of the form:
\begin{equation}
\psi(q_1,q_2,\dots,q_N) = \mathcal{N}\exp\left[\frac{i}{8\pi}(A_1q_1 + A_2q_2 +\cdots + A_Nq_N)\right].
\end{equation}
Hence the wave functions are no more functionals, but ordinary functions of a finite number of variables $q_1, q_2,\dots,q_N$. The inner product between the wave functions is defined as:
\begin{equation} 
\langle\psi_1\vert\psi_2\rangle := \int_{-L/2}^{L/2}dq_1\int_{-L/2}^{L/2}dq_2\cdots\int_{-L/2}^{L/2}dq_N\,\psi^*_1(q_1,q_2,\dots,q_N)\psi_2(q_1,q_2,\dots,q_N).
\end{equation}
and if we want the wave function $\psi(q_1,q_2,\dots,q_N)$ to be normed to unity, we may take:
\begin{equation}
\mathcal{N} = \frac{1}{\sqrt{L^N}}.
\end{equation}
So we find:
\begin{equation}
\psi(q_1,q_2,\dots,q_N) = \frac{1}{\sqrt{L^N}}\exp\left[\frac{i}{8\pi}(A_1q_1 + A_2q_2 +\cdots + A_Nq_N)\right].
\end{equation}
Eqs. (3.5) and (4.29) imply that the energy eigenvalues of the gravitational field from the point of view of our accelerating observer are:
\begin{equation}
E = \frac{\alpha a}{8\pi}(n_1 + n_2 +\cdots n_N),
\end{equation}
where 
\begin{equation}
\alpha := \frac{16\pi^2}{L}.
\end{equation}

   As the reader may have noticed, quantization of the gravitational field in flat spacetime from the point of view of an accelerating observer is, in our model, very similar to the quantization of the system of {\it free particles} in non-relativistic quantum mechanics. In such systems the coordinates of the momentum space are constants of motion, and the coordinates of the configuration space are, possibly up to additive constants, linear functions of time, whereas in our system the momentum variables $p(\chi)$ are constants , and the corresponding coordinates $q(\chi)$ of the configuration space depend, up to additive constants, linearly on the proper time $\tau$ of the observer. When quantizing the system of free particles in non-relativistic quantum mechanics one must use {\it box normalization} for the wave function of the system. When using the box normalization one assumes that the particles of the system lie in a rectangular box. Finally, at the end of the calculations, the edge kength of the box is taken to infinity. In our model we also applied a sort of box normalization, where the coordinates of the confuguration space of the system were assumed to lie within the interval $[-L/2,L/2]$. However, we did not take the "edge length" $L$ of the box to infinity, but we took $L$ to be a {\it fixed number}. As a consequence, $L$ appears as a {\it parameter} of the model, which must be fixed such that the predictions of the model agree with experiments.
  
\section{Unruh Effect}

In Section 4 we completed, in our model, the quantization of gravity in flat spacetime. The most important outcome of our model was that the area of an accelerating plane has a discrete spectrum with equal spacing. We are now prepared to go to the thermodynamical properties of gravity. As it is well known, the thermodynamical properties of any system may be deduced from its partition function
\begin{equation}
Z(\beta) := \sum_n e^{-\beta E_n},
\end{equation}
where $\beta$ is the temperature parameter, and we have summed over the energy eigenstates $n$ of the system. In what follows, we shall base our calculation of the partition function on Eq. (4.34). We say that constituent $j$ is in {\it vacuum}, if $n_j=0$; otherwise the constituent is in an {\it excited state}. 

  When obtaining an expression for the partition function of our system we shall assume that at least one of the constituents of the accelerated plane is in an excited state, and when the quantum states of the constituents are interchanged, the quantum state of the system will also change. With these assumptions the partition function takes the form:
\begin{equation}
\begin{split}
Z(\beta) = &\,\,\,\,\sum_{n_1=1}^\infty\exp\left(-\frac{\alpha\beta a}{8\pi}n_1\right)\\
                &+\sum_{n_1=1}^\infty\exp\left(-\frac{\alpha\beta a}{8\pi}n_1\right)\sum_{n_2=1}^\infty\exp\left(-\frac{\alpha\beta a}{8\pi}n_2\right)\\
               &+\cdots\\
               &+\sum_{n_1=1}^\infty\exp\left(-\frac{\alpha\beta a}{8\pi}n_1\right)\cdots\sum_{n_N=1}^\infty\exp\left(-\frac{\alpha\beta a}{8\pi}n_N\right).
\end{split}
\end{equation}
In the first term on the right hand side of Eq. (5.2)  just one of the constituents is in an excited state, and we have summed over all of those states. In the second term two of the constituents are in excited states. Finally, in the last term all of the $N$ constituents are in excited states. Defining the {\it characteristic temperature}
\begin{equation}
T_C := \frac{\alpha a}{8\pi\ln(2)}
\end{equation}
we may write Eq. (5.2) as:
\begin{equation}
\begin{split}
Z(\beta) &= \frac{1}{2^{\beta T_C} - 1} + \left(\frac{1}{2^{\beta T_C} - 1}\right)^2 +\cdots + \left(\frac{1}{2^{\beta T_C} - 1}\right)^N\\
             &= \frac{1}{2^{\beta T_C} - 2}\left[1 - \left(\frac{1}{2^{\beta T_C} - 1}\right)^N\right],
\end{split}
\end{equation}
whenever $\beta T_C\ne 1$. When $\beta T_C=1$, we have:
\begin{equation}
Z(\beta) = N.
\end{equation}

    The energy of the gravitational field may be calculated from its partition function $Z(\beta)$ as:
\begin{equation}
E = -\frac{\partial}{\partial\beta}\ln[Z(\beta)],
\end{equation}
and if we define the energy per constituent as 
\begin{equation}
\bar{E} := \frac{E}{N},
\end{equation}
Eqs. (5.4) and (5.6) imply:
\begin{equation}
\bar{E} = \bar{E}_1 + \bar{E}_2,
\end{equation}
where we have denoted:
\begin{subequations}
\begin{eqnarray}
\bar{E}_1 &:=& \frac{1}{N}\frac{2^{\beta T_C}}{2^{\beta T_C} - 2}T_C\ln(2),\\
\bar{E}_2 &:=& \frac{2^{\beta T_C}}{2^{\beta T_C} - 1 -(2^{\beta T_C} - 1)^{N+1}}T_C\ln(2),
\end{eqnarray}
\end{subequations}
whenever $\beta T_C\ne 1$. When $\beta T_C=1$, a calculation similar to the one performed in Ref. \cite{seite} gives, in the large $N$ limit, the result: 
\begin{equation}
\bar{E} = T_C\ln(2).
\end{equation}
Eqs. (4.34), (5.3), (5.8) and (5.9) imply that the average
\begin{equation}
\bar{n} := \frac{n_1 + n_2 +\cdots n_N}{N}
\end{equation}
of the quantum numbers $n_1, n_2,\dots, n_N$ depends of the temperature parameter $\beta$ as:
\begin{equation}
\bar{n} = \frac{1}{N}\frac{2^{\beta T_C}}{2^{\beta T_C} - 2} + \frac{2^{\beta T_C}}{2^{\beta T_C} - 1 - (2^{\beta T_C} - 1)^{N+1}}.
\end{equation}

   So far everything has been quite exact, and no approximations have been made. At this point we proceed to consider, what happens in the limit, where the number $N$ of the constituents of the accelerating plane tends to infinity. In this limit the first term on the right hand side of Eq. (5.12) will vanish, and we have, in effect:
\begin{equation}
\bar{n} = \frac{2^{\beta T_C}}{2^{\beta T_C} - 1 - (2^{\beta T_C} - 1)^{N+1}}.
\end{equation}
When $\beta T_C >1$, which means that the temperature $T$ experienced by the observer moving along the plane is less than the characteristic temperature $T_C$ we have:
\begin{equation}
\lim_{N\rightarrow\infty}(2^{\beta T_C} - 1)^{N+1} = \infty,
\end{equation}
and so the constituents of the plane are, in effect, in vacuum. However, when $\beta T_C<1$, which means that $T>T_C$, we find:
\begin{equation}
\lim_{N\rightarrow\infty}(2^{\beta T_C} - 1)^{N+1} = 0,
\end{equation}
and so we may write, as an excellent approximation:
\begin{equation}
\bar{n} = \frac{2^{\beta T_C}}{2^{\beta T_C} - 1}.
\end{equation}
An interesting property of this expression is that, when $T$ tends to $T_C$ from the right hand side, which means that $\beta T_C$ tends to $1$ from the left hand side, we have:
\begin{equation}
\bar{n} = 2.
\end{equation}
So we find that at the characteristic temperature $T_C$ the accelerating plane performs a {\it phase transition}, where the constituents of the plane jump, in average, from the vacuum to the second excited states. The latent heat per constituent associated with this phase transition is 
\begin{equation}
\bar{L} = 2T_C\ln(2).
\end{equation}

   Since the constituents of the plane are effectively in the vacuum, when $T<T_C$, and jump to the second excited states at the characteristic temperature $T_C$, we may view the characteristic temperature $T_C$, defined in Eq. (5.3), as the {\it lowest possible temperature} experienced by the accelerating observer. Interestingly, we find that if we choose:
\begin{equation}
\alpha = 4\ln(2),
\end{equation}
then:
\begin{equation}
T_C = \frac{a}{2\pi},
\end{equation}
which exactly agrees with the {\it Unruh temperature} \cite{viitoo}
\begin{equation}
T_U := \frac{a}{2\pi}
\end{equation}
experienced by the observer. In this sense we have been able to derive the {\it Unruh effect} from our model of quantum gravity in flat spacetime: An accelerating observer detects thermal radiation with a temperature, which is proportional to the proper acceleration $a$ of the observer. That radiation is produced, when the constituents of the plane perform quantum jump from the excited states to the vacuum.  

\section{Concluding Remarks}

  In this paper we have constructed, step by step, a model of quantum gravity in flat spacetime from the point of view of a unformly accelerating observer. An observer of this kind detects phenomena identical to those in a uniform gravitational field in his surroundings, and we considered those effects quantum-mechanically. Among other things, our nodel of quantum gravity predicted that spacetime has a discrete structure in the sense that the plane at rest with respect to the accelerating observer consists of separate constituents, each of which has an area, which is an integer times a certain fundamental area. As a consequence, the area eigenvalues of the accelerating plane are of the form:
\begin{equation}
A = \alpha(n_1 + n_2 +\cdots n_N),
\end{equation}
where the quantum numbers $n_j$ are non-negative integers for all $j = 1,2,\dots, N$, and $N$ is the number of the constituents of the plane. We found that with the choice 
\begin{equation}
\alpha = 4\ln(2)
\end{equation}
for the parameter $\alpha$ of our model an observer with constant proper acceleration $a$ detects thermal radiation with certain minimum temperature, which agrees with the {\it Unruh temperature}
\begin{equation}
T_U = \frac{a}{2\pi}
\end{equation}
of the observer. In this sense our model of quantum gravity in flat spacetime implies the Unruh effect.

   Basically, the results of the paper were outcomes of our decision to supplement the standard Einstein-Hilbert action with the standard Gibbons-Hawking boundary term. We first picked up a rectangular box on a spacelike hypersurface of the flat Minkowski spacetime spacetime,  where the time coordinate is a constant, and then put one of its faces into a uniformly accelerating motion, while the other faces were kept at rest in the flat Minkowski system of coordinates. When the faces of the box proceeded in the flat spacetime, a three-dimensional boundary was created for spacetime. It turned out that the trace of the exterior curvature tensor on the three-dimensional timelike hypersurface created by the accelerating plane was non-zero, while the exterior curvature tensor on the timelike hypersurfaces created by the other faces of the box vanished identically. As a consequence, the Gibbons-Hawking boundary term was non-zero, even though the Einstein-Hilbert action vanished. Thus, the presence of the Gibbons-Hawking boundary term produced non-zero gravitational action, despite of the fact that spacetime was flat. 

    From the action we obtained the classical Hamiltonian with respect to the accelerating observer. The discrete structure of the accelerating plane followed from the straightforward replacement of the classical Hamiltonian by the corresponding Hamiltonian operator according to the standard rules of quantum mechanics. The coordinates $p_j$ of the momentum space agree, up to the coefficient $\frac{1}{8\pi}$, with the areas of the constituents of the plane, whereas the corresponding coordinates $q_j$ of the configuration space agree, up to additive constants, with the boost angle of the observer. In the process we were compelled to introduce a box normalization for the wave function of the gravitational field, where the coordinates $q_j$ of the configuration space are confined to live inside of a box with edge length $L$. {\footnote{Not to be confused with the edge length of the accelerating box in flat spacetime!}} Defining a parameter $\alpha$ in terms of the edge length $L$ as:
\begin{equation}
\alpha = \frac{16\pi^2}{L}
\end{equation}
we obtained the area spectrum in Eq. (6.1).

  As we have seen, quantization of gravity in flat spacetime from the point of view of an accelerating observer is really very simple and straightforward, and may be carried out explicitly. Nevertheless, there are still some shortcomings in our procedure. For instance, our model of quantum gravity predicted for the areas of the constituents of the accelerating plane both positive and {\it negative} eigenvalues, and we excluded the negative eigenvalues "by hand". An even more serious problem in our approach is the presence of an undetermined parameter, the edge length $L$ of the box used in the normalization of the wave function. Eqs. (6.2) and (6.4) imply that if we choose 
\begin{equation}
L = \frac{4\pi^2}{\ln(2)}\approx 57.0,
\end{equation}
our model predicts the Unruh effect. The close relationship between the variables $q_j$ and  the boost angle $\phi$ of the observer suggests that the edge length $L$ sets the bound for $\phi$ such that, for symmetry reasons, $-\frac{L}{2}\le\phi\le\frac{L}{2}$. Whether such interpetation is valid, is still an open question, as well as are its possible consequences. 

     Taken as a whole, our paper highlights an aspect of the potential quantum theory of gravity, which so far has received very little attention: The quantum states of the gravitational field are not absolute, but {\it relative}, and they depend on the  state of motion of the observer. For instance, an accelerating observer in flat spacetime detects gravity-like effects, which may be quantized by means of the standard rules of quantum mechanics, whereas inertial observers detect no effects of gravity whatsoever. It is to be hoped that the results obtained this paper would pave the way for a proper quantum theory of gravitation in curved spacetime.


\begin{thebibliography} {20}
\bibitem{vii} There are still conributions to the boundary  term emerging from the edges of our box. Those contributions, however, may be cancelled, if necessary, by means of an addition of an appropriate, constant boundary term.

\medskip
\end{thebibliography}
\end{document}